\documentclass[12pt]{article}
\usepackage{amsmath,epsfig,euscript,bm,url}

\usepackage{graphicx,epstopdf}

\newcommand{\gzp}{g_{z^\prime}}

\begin{document}

\begin{titlepage}

\begin{flushright}
WSU-HEP-1714\\
November 7, 2017\\
\end{flushright}

\vspace{0.7cm}
\begin{center}
\Large\bf\boldmath
Remarks on the $\bm{Z^\prime}$ Drell-Yan cross section
\unboldmath
\end{center}

\vspace{0.8cm}
\begin{center}
{\sc Gil Paz and Joydeep Roy}\\
\vspace{0.4cm}
{\it 
Department of Physics and Astronomy \\
Wayne State University, Detroit, Michigan 48201, USA 
}

\end{center}
\vspace{1.0cm}
\begin{abstract}
  \vspace{0.2cm}
  \noindent
Many extensions of the standard model contain an extra $U(1)'$ gauge group with a heavy $Z^\prime$ gauge boson. Perhaps the most clear signal for such a $Z^\prime$ would be a resonance in the invariant mass spectrum of the lepton pairs to which it decays. In the absence of such a signal, experiments can set limits on the couplings of such a $Z^\prime$, using a standard formula from theory. We repeat its derivation and find that, unfortunately, the standard formula in the literature is a factor of 8 too small. We briefly explore the implication for existing experimental searches and encourage the high energy physics community to re-examine analyses that have used this formula. 
\end{abstract}
\vfil

\end{titlepage}

\section{Introduction}
Many models of physics beyond the standard model (BSM) include an extra $U(1)'$ gauge group with a heavy $Z^\prime$ gauge boson, see for example the review \cite{Langacker:2008yv}. Perhaps the most clear signal for such a $Z^\prime$ would be a resonance in the invariant mass spectrum of the lepton pairs to which it decays. So far, no such $Z^\prime$ has been observed. See, for example, the recent ATLAS \cite{Aaboud:2017buh} and CMS \cite{Khachatryan:2016zqb} searches. Such experimental constraints on a $Z^\prime$ can be very useful in BSM model building, \emph{provided} that they are presented in an appropriate form. In practice, experiments usually report their data in three ways.   

The first is reporting limits on the $Z^\prime$/$Z$ cross section ratio to reduce experimental uncertainties as is done, e.g.,  in \cite{Khachatryan:2016zqb} and \cite{Khachatryan:2014fba}.  These are not so easy to use, since they require the knowledge of the Standard Model (SM) $Z$ cross section at NNLO to convert to limits  on  $Z^\prime$. Furthermore, results are currently presented only as figures without the detailed numerical values. 

The second is to use some ``benchmark" models such as the sequential standard model and $E_6$ models, see  \cite{Langacker:2008yv} for a review. Having fixed the coupling of the $Z^\prime$, experiments can report limits on the mass of these specific $Z^\prime$ gauge bosons, e.g. \cite{Aaboud:2017buh}. These are useful in tracking the progress of the experimental constraints over time, but are not easily translatable to other $Z^\prime$ models. 

The most useful approach was presented in \cite{Carena:2004xs}. For a given $Z^\prime$ mass and  center of momentum energy, the cross section $\sigma(h_1+h_2 \to Z^\prime+X\to \ell^+\ell^{-}\!+X)$ is expressed as a sum over products of two quantities. One, $\omega_i$, depends on SM Parton Distribution Functions (PDFs) and is independent of the $Z^\prime$ model, and the other, $c_i$, depends on the couplings of the $Z^\prime$ and is model dependent. See the next section for exact definitions. Experiments obtain constraints on the cross section times the branching ratio and they can translate them to constraints on the parameters $c_i$. See, for example, a recent exclusion plot in \cite{Khachatryan:2014fba}.  The formula given in \cite{Carena:2004xs} is 
\begin{equation}\label{sigma48}
\sigma(h_1+h_2 \to Z^\prime+X\to \ell^+\ell^{-}\!+X)=\dfrac{\pi}{48s}\left[c_uw_u(s,M^2_{Z^\prime})+c_dw_d(s,M^2_{Z^\prime})\right].
\end{equation}
We have re-derived this formula and we find that it is too small by a factor of 8, i.e. the denominator should be $6s$. The purpose of this paper is to briefly explore the implications on the $Z^\prime$ experimental  limits and bring the issue to the attention of the high energy physics community. 

The rest of the paper is structured as follows. In section \ref{theory} we re-derive the expression for the cross section and compare it to the expressions in references  \cite{Hamberg:1990np} and \cite{Carena:2004xs}. In section \ref{experiment} we explore the implications on the existing experimental constraints. We present our conclusions in section \ref{conclusions}.  

\section{Theoretical expressions}\label{theory}

\subsection{$\bm{Z^\prime}$ Drell-Yan cross section at leading order}
Consider a $Z^\prime$ flavor-diagonal coupling of the form \cite{Langacker:2008yv} 
\begin{equation}\label{Lcoupling}
{\cal L}=-\gzp Z^\prime_\mu\sum_i\bar f_i \gamma^\mu\left[\epsilon_L^iP_L+\epsilon_R^iP_R\right]f_i=-\dfrac{\gzp }{2}Z^\prime_\mu\sum_i\bar f_i \gamma^\mu\left[g_V^i-g_A^i\gamma^5\right]f_i,
\end{equation} 
where $P_{L,R}=\left(1\mp\gamma^5\right)/2$ and $g_{V,A}^i=\epsilon_L^i\pm\epsilon_R^i$. Averaging over the colors of the initial quarks, the cross section for the process $q\bar q\to f_i\bar f_i$ is \cite{Griffiths:2008zz} 
\begin{equation}
\sigma(q\bar q\to Z^\prime\to f_i\bar f_i)=\dfrac{\gzp^4}{\left(Q^2-M^2_{Z^\prime}\right)^2+\Gamma_{Z^\prime}^2M_{Z^\prime}^2}\dfrac1{N_c}\dfrac1{3\pi}\dfrac{Q^2}{64}\Big[(g_V^q)^2+(g_A^q)^2\Big]\Big[(g_V^f)^2+(g_A^f)^2\Big],
\end{equation}
where $Q^2$ is the invariant  mass of the $f_i-\bar f_i$ pair,  $M_{Z^\prime}$ is the $Z^\prime$ mass, $\Gamma_{Z^\prime}$ is the $Z^\prime$ width, and $N_c=3$.

At leading order in $\alpha_s$ the Drell-Yan cross section is given by \cite{Patrignani:2016xqp} 
\begin{eqnarray}
&&\sigma(h_1+h_2 \to Z^\prime+X\to f_i \bar f_i+X)=\nonumber\\
&=&\sum_q\int_0^1 dx_1 \int_0^1 dx_2\, \left[f_{q/h_1}(x_1) f_{\bar q/h_2}(x_2)+\left(x_1\leftrightarrow x_2, h_1\leftrightarrow h_2\right)\right]\sigma_{q\bar q\to Z^\prime\to f_i\bar f_i }.
\end{eqnarray}

To find an expression for $d\sigma(h_1+h_2 \to Z^\prime+X)/dQ^2$, we insert unity as $1=\int dQ^2\delta(Q^2-x_1x_2s)$ and take a derivative with respect to $Q^2$. Here, $s=(p_1+p_2)^2$, where $p_i$ is the four-momentum of $h_i$. We get 
\begin{eqnarray}\label{dsigmaQ2}
&&\dfrac{d\sigma(h_1+h_2 \to Z^\prime+X\to f_i \bar f_i+X)}{dQ^2}=\nonumber\\
&=&\sum_q\int_0^1 dx_1 \int_0^1 dx_2\, \left[f_{q/h_1}(x_1) f_{\bar q/h_2}(x_2)+\left(x_1\leftrightarrow x_2, h_1\leftrightarrow h_2\right)\right]\,\sigma_{q\bar q\to Z^\prime\to f_i\bar f_i }\delta(Q^2-x_1x_2s)\nonumber\\
&=&\sum_q\int_0^1 dx_1 \int_0^1 dx_2\, \left[f_{q/h_1}(x_1) f_{\bar q/h_2}(x_2)+\left(x_1\leftrightarrow x_2, h_1\leftrightarrow h_2\right)\right]\delta\left(1-\frac{x_1x_2s}{Q^2}\right)\times\nonumber\\
&\times&\dfrac{\gzp^4}{\left(Q^2-M^2_{Z^\prime}\right)^2+\Gamma_{Z^\prime}^2M_{Z^\prime}^2}\dfrac1{N_c}\dfrac1{3\pi}\dfrac{1}{64}\Big[(g_V^q)^2+(g_A^q)^2\Big]\Big[(g_V^f)^2+(g_A^f)^2\Big].
\end{eqnarray}  

As a first check,  we note that for a photon, $M_{Z^\prime}=0$, $\Gamma_{Z^\prime}=0$, $\gzp=e$, $g_A^i=0$, and $g_V^i=2e_i$, where $e_i$ is the fermion electric charge in units of the positron charge.  For an $l^+l^-$ final state equation (\ref{dsigmaQ2}) becomes 
\begin{eqnarray}\label{gamma}
&&\dfrac{d\sigma(h_1+h_2 \to l^+l^-+X)}{dQ^2}=\\
&=&\dfrac{4\pi\alpha^2}{9Q^4}\sum_qe_q^2\int_0^1 dx_1 \int_0^1 dx_2\, \left[f_{q/h_1}(x_1) f_{\bar q/h_2}(x_2)+\left(x_1\leftrightarrow x_2, h_1\leftrightarrow h_2\right)\right]
\delta\left(1-\frac{x_1x_2s}{Q^2}\right),\nonumber
\end{eqnarray}  
which is a well-known result \cite{Halzen:1984mc}. 

As a second check, we consider the case of the SM $Z$. We have \cite{Patrignani:2016xqp} $g_V^i=t^i_{3L}-2e_i\sin^2\theta_W$,  $g_A^i=t^i_{3L}$, and $\gzp=e/(\sin\theta_W\cos\theta_W)$, where $t^i_{3L}$ is the weak isospin of fermion $i$ ($+1/2$ for up-type quark and neutrino,  $-1/2$ for down-type quark and a charged lepton),  and $\theta_W$ is the weak angle.  For $Z$ decay to $\ell^+\ell^-$ equation (\ref{dsigmaQ2}) becomes 
\begin{eqnarray}\label{Z}
&&\dfrac{d\sigma(h_1+h_2 \to Z+X\to f_i \bar f_i+X)}{dQ^2}\\
&=&\sum_q\int_0^1 dx_1 \int_0^1 dx_2\, \left[f_{q/h_1}(x_1) f_{\bar q/h_2}(x_2)+\left(x_1\leftrightarrow x_2, h_1\leftrightarrow h_2\right)\right]\delta\left(1-\frac{x_1x_2s}{Q^2}\right)\nonumber\\
&\times&\dfrac{\pi\alpha^2}{\sin^4\theta_W\cos^4\theta_W}\dfrac1{\left(Q^2-M^2_{Z}\right)^2+\Gamma_{Z}^2M_{Z}^2}\dfrac1{3N_c}\dfrac{1}{64}\Big[4\left(t^q_{3L}-2e_q\sin^2\theta_W\right)^2+4\left(t^q_{3L}\right)^2\Big]\Big[1+(1-4\sin^2\theta_W)^2\Big].\nonumber
\end{eqnarray}  
We will show in section \ref{sec:Hamberg} that this result agrees with  \cite{Hamberg:1990np}. 

Having performed the checks for a photon and the SM $Z$, we return to the general $Z^\prime$ case. In the narrow width approximation the expression for the cross section, equation (\ref{dsigmaQ2}), can be simplified by using  
\begin{equation}
\dfrac1{\left(Q^2-M^2_{Z^\prime}\right)^2+\Gamma_{Z^\prime}^2M_{Z^\prime}^2}\to\dfrac{\pi}{\Gamma_{Z^\prime}M_{Z^\prime}}\delta\left(Q^2-M_{Z^\prime}^2\right),
\end{equation}
and \cite{Halzen:1984mc}
\begin{equation}
\mbox{BR}\left(Z^\prime\to \ell^+\ell^-\right)=\dfrac{\Gamma_{Z^\prime}(Z^\prime\to \ell^+\ell^- )}{\Gamma_{Z^\prime}}=\dfrac{\gzp^2\Big[(g_V^\ell)^2+(g_A^\ell)^2\Big]M_{Z^\prime}}{48\pi\,\Gamma_{Z^\prime}}.
\end{equation}
Setting $N_c=3$, we obtain
\begin{equation}\label{sigma6}
\sigma(h_1+h_2 \to Z^\prime+X\to \ell^+\ell^{-}\!+X)=\dfrac{\pi}{6s}\left[c_uw_u(s,M^2_{Z^\prime})+c_dw_d(s,M^2_{Z^\prime})\right],
\end{equation}
where 
\begin{equation}
c_{u,d}=\dfrac12\gzp^2\left[(g_V^{u,d})^2+(g_A^{u,d})^2\right]\mbox{BR}\left(Z^\prime\to \ell^+\ell^-\right),
\end{equation}
and at leading order in $\alpha_s$
\begin{equation}
w_q(s,M^2_{Z^\prime})=\int_0^1 dx_1 \int_0^1 dx_2\, \left[f_{q/h_1}(x_1) f_{\bar q/h_2}(x_2)+\left(x_1\leftrightarrow x_2, h_1\leftrightarrow h_2\right)\right]\delta\left(\dfrac{M^2_{Z^\prime}}{s}-x_1x_2\right).
\end{equation}

Equation (\ref{sigma6}) is the main result of this paper. Notice the factor of 6 in the denominator compared to equation (\ref{sigma48}).

\subsection{Comparison to  reference \cite{Hamberg:1990np}}\label{sec:Hamberg}
To check our calculation, we compare it to \cite{Hamberg:1990np}. In that paper the $\alpha_s^{2}$ correction to the Drell-Yan $K$ factor were calculated. We will only need the  $\alpha_s^0$ terms. Equation (2.22) of \cite{Hamberg:1990np} for a vector boson-quark coupling is 
\begin{equation}
Vq_i\bar q_j: \quad ig_V\gamma_\mu(v_i^V+a_i^V\gamma_5). 
\end{equation}

The first two lines of equation (A.13) of \cite{Hamberg:1990np} are 
\begin{eqnarray}
v_u^\gamma=\dfrac23,&\qquad& a_u^\gamma=0\nonumber\\
v_d^\gamma=-\dfrac13,&\qquad& a_d^\gamma=0. 
\end{eqnarray}
Comparing to equation (\ref{Lcoupling}), we get that for a photon coupling $g_\gamma=-e$.

The last two lines of equation (A.13) of \cite{Hamberg:1990np} are 
\begin{eqnarray}
v_u^Z=1-\dfrac83\sin^2\theta_W,&\qquad& a_u^Z=-1\nonumber\\
v_d^Z=-1+\dfrac43\sin^2\theta_W,&\qquad& a_d^Z=1.
\end{eqnarray}
Comparing to equation (\ref{Lcoupling}), we get that for a $Z$ coupling 
\begin{equation}\label{Hamberg_gz}
\gzp=-\dfrac{e}{\bm{4}\sin\theta_W\cos\theta_W}.
\end{equation}
Notice the factor of 4 compared to the standard expression \cite{Patrignani:2016xqp} . It implies that one should be careful in adapting the results of \cite{Hamberg:1990np} to the case of a $Z^\prime$.  Equation (2.2) of  \cite{Hamberg:1990np} is 
\begin{equation}\label{Hamberg}
\dfrac{d\sigma_V}{dQ^2}=\tau\sigma_V(Q^2,M_V^2)W_V(\tau,Q^2),\qquad \tau=\dfrac{Q^2}{s}.  
\end{equation}
The ${\cal O}(\alpha_s^0)$ expression for $W_{\gamma,Z}(\tau,Q^2)$ from equation (A.20) of  \cite{Hamberg:1990np}  is
\begin{equation}\label{Wv}
W_{\gamma,Z}(\tau,Q^2)=\int_0^1 dx_1 \int_0^1 dx_2\,\delta(\tau-x_1x_2)\sum_{i,j\in Q,\bar Q} \delta_{ij}(v_i^2+a_i^2)q_i(x_1)\bar q_j(x_2),  
\end{equation}
where $q_i$ ($\bar q_j$) is the quark (anti-quark) PDFs.  

Reference \cite{Hamberg:1990np} calls $\sigma_V$ the ``pointlike cross section", although it is  the pointlike $d\sigma_V/dQ^2$. Equation (A.1) of \cite{Hamberg:1990np} is 
\begin{equation}
\sigma_\gamma(Q^2)=\dfrac{4\pi\alpha^2}{3Q^4}\dfrac1{N_c}.
\end{equation}
Combining it with (\ref{Wv}) in (\ref{Hamberg}), we get our equation (\ref{gamma}).  

Combining equation (A.2) and (A.5) of \cite{Hamberg:1990np}  we have 
\begin{equation}
\sigma_Z(Q^2,M_Z^2)=\dfrac{\pi\alpha}{4M_Z\sin^2\theta_W\cos^2\theta_W}\dfrac1{N_c}\dfrac1{\left(Q^2-M^2_{Z^\prime}\right)^2+\Gamma_{Z^\prime}^2M_{Z^\prime}^2}\dfrac{\alpha M_z\Big[1+(1-4\sin^2\theta_W)^2\Big]}{48\sin^2\theta_W\cos^2\theta_W}.
\end{equation}
Combining this with (\ref{Wv}) in (\ref{Hamberg}), we get our equation (\ref{Z}). 

In summary we find agreement between our expressions and  \cite{Hamberg:1990np}  for the ${\cal O}(\alpha_s^0)$ expression for $\gamma$  and $Z$ Drell-Yan processes. 

\subsection{Comparison to reference \cite{Carena:2004xs}}
We now compare our results to reference \cite{Carena:2004xs}, that first presented (\ref{sigma48}). Equation (2.8) of \cite{Carena:2004xs} is
\begin{equation}\label{Ccoupling}
\sum_fz_fg_zZ^\prime_\mu\bar f\gamma^\mu f,
\end{equation}
``...where $f=e^j_R, l_L^j , u^j_R, d^j_R, q_L^j $ are the usual lepton and quark fields in the weak eigenstate basis; $l_L^j = (\nu_L^j , e^j_L)$ and $q_L^j = (u^j_L, d^j_L)$ are the $SU(2)_W$ doublet fermions. The index $j$ labels the three fermion generations. Altogether there are 15 fermion charges, $z_f$ ." \cite{Carena:2004xs}

Equation (3.1) of  \cite{Carena:2004xs} is
\begin{equation}
\dfrac{d\sigma(p\bar p \to Z^\prime+X\to l^+l^-+X)}{dQ^2}=\dfrac1{s}\sigma(Z^\prime\to l^+l^-)W_{Z^\prime}(s,Q^2),
\end{equation}
where we  ignore  the interference of the $Z^\prime$ with the $Z$ and the photon. According to equation (3.2) of \cite{Carena:2004xs} 
\begin{equation}
\sigma(Z^\prime\to l^+l^-)=\dfrac{g_z^2}{4\pi}\left(\dfrac{z_{l_j}^2+z^2_{e_j}}{288}\right)\dfrac{Q^2}{\left(Q^2-M^2_{Z^\prime}\right)^2+\Gamma_{Z^\prime}^2M_{Z^\prime}^2}.
\end{equation}
At ${\cal O}(\alpha_s^0)$ and for generation independent $Z'$ coupling 
\begin{equation}
W_{Z^\prime}(s,Q^2)
=g_z^2\left[(z_q^2+z^2_u)w_u(s,Q^2)+(z_q^2+z^2_d)w_u(s,Q^2)\right],
\end{equation}
where 
\begin{equation}
w_{u(d)}=\sum_{q=u,c,(d,s,b)}\int_0^1 dx_1 \int_0^1 dx_2\,\left[ f_{q/P}(x_1) f_{\bar q/\bar P}(x_2)+\left(x_1\leftrightarrow x_2, P\leftrightarrow \bar P\right)\right]\delta\left(\dfrac{Q^2}{s}-x_1x_2\right),
\end{equation}
and we ignore the scale dependance of the PDFs at ${\cal O}(\alpha_s^0)$. 

Comparing  (\ref{Ccoupling}) to  (\ref{Lcoupling}), we have $g_z=-\gzp$, and $z_{l_j}=z_{e^j_L}=(g_V^{e_j}+g_A^{e_j})/2$, $z_{e_j}=z_{e_R^j}=(g_V^{e_j}-g_A^{e_j})/2$, $z_{u_L^j}=z_{q}=(g_V^{u}+g_A^{u})/2$, $z_{d_L^j}=z_{q}=(g_V^{d}+g_A^{d})/2$, $z_{u,d}=(g_V^{u,d}-g_A^{u,d})/2$. As a result, see equation (\ref{Lcoupling}), 
\begin{eqnarray}
z_{l_j}^2+z^2_{e_j}&=&\dfrac{(g_V^{e_j}+g_A^{e_j})^2}{4}+\dfrac{(g_V^{e_j}-g_A^{e_j})^2}{4}=\dfrac{(g_V^{e_j})^2+(g_A^{e_j})^2}{2}=(\epsilon_L^{e_j})^2+(\epsilon_R^{e_j})^2\nonumber\\
z_{q}^2+z^2_{u}&=&\dfrac{(g_V^{u}+g_A^{u})^2}{4}+\dfrac{(g_V^{u}-g_A^{u})^2}{4}=\dfrac{(g_V^{u})^2+(g_A^{u})^2}{2}=(\epsilon_L^{u})^2+(\epsilon_R^{u})^2\nonumber\\
z_{q}^2+z^2_{d}&=&\dfrac{(g_V^{d}+g_A^{d})^2}{4}+\dfrac{(g_V^{d}-g_A^{u})^2}{4}=\dfrac{(g_V^{d})^2+(g_A^{d})^2}{2}=(\epsilon_L^{d})^2+(\epsilon_R^{d})^2.
\end{eqnarray}

Altogether we find 
\begin{eqnarray}
&&\dfrac{d\sigma(p\bar p \to Z^\prime+X\to l^+l^-+X)}{dQ^2}=\sum_{q=u,d}\int_0^1 dx_1 \int_0^1 dx_2\,\left[ f_{q/P}(x_1) f_{\bar q/\bar P}(x_2)+\left(x_1\leftrightarrow x_2, P\leftrightarrow \bar P\right)\right]\times\nonumber\\
&\times&\delta\left(1-\frac{x_1x_2s}{Q^2}\right)\times\dfrac{g_z^4}{\left(Q^2-M^2_{Z^\prime}\right)^2+\Gamma_{Z^\prime}^2M_{Z^\prime}^2}\dfrac1{N_c}\dfrac1{3\pi}\dfrac{1}{\bm{512}}\Big[(g_V^q)^2+(g_A^q)^2\Big]\Big[(g_V^e)^2+(g_A^e)^2\Big],
\end{eqnarray}
In other words, the result of  \cite{Carena:2004xs} is 8 times too small compared to  our equation (\ref{Z}). 

Using the narrow width approximation \cite{Carena:2004xs} obtained
\begin{equation}
\sigma(h_1+h_2 \to Z^\prime+X\to \ell^+\ell^{-}\!+X)=\dfrac{\pi}{48s}\left[c_uw_u(s,M^2_{Z^\prime})+c_dw_d(s,M^2_{Z^\prime})\right].
\end{equation}
Again this expression is 8 times too small compared to our equation (\ref{sigma6}). The wrong equation (\ref{sigma48}) also appears in the ``$Z^\prime$-boson searches" review by the PDG \cite{Patrignani:2016xqp},  and in \cite{Accomando:2010fz} often cited together with \cite{Carena:2004xs} by ATLAS and CMS papers on $Z^\prime$ searches.

Interestingly, there are two papers, \cite{Feldman:2006wb} and \cite{deBlas:2012qp},  that have the correct numerical factor but do not note  the discrepancy.  In general, the expression for the cross section is meaningful only if one also defines $c_{u,d}$ in terms of the $Z^\prime$ charges and $w_{u,d}$ in terms of the PDFs. Furthermore, one has to define the $Z^\prime$ charges by presenting the Lagrangian since, unlike the PDFs, there is more than one convention for them in the literature.  These three conditions were met in \cite{Feldman:2006wb}. Equation (6.5) of the published version of that paper has the right numerical factor, but the authors state in a footnote ``We note that the analysis of ref. [18] absorbs a factor of 8 in their PDFs contained within the function, defined as $W_{Z^\prime}$"  \cite{Feldman:2006wb}. We checked the definition of $w_{u,d}$ in  \cite{Feldman:2006wb}, where it is called ${\cal W}_{\{A B (q\bar q)\}}$, and it is the same\footnote{Daniel Feldman informed us \cite{Feldman:2017} that they meant to imply that \emph{numerically}  \cite{Carena:2004xs} appeared to include the factor of 8, e.g. in the figures of  \cite{Carena:2004xs}, but it was missing in the \emph{analytical} expression of  \cite{Carena:2004xs}.} as the ${\cal O}(\alpha_s^0)$ expression for $w_{u,d}$ in  \cite{Carena:2004xs}. Only two of the conditions were met in \cite{deBlas:2012qp} that defines $c_{u,d}$ and the $Z^\prime$ charges, but not $w_{u,d}$ for which they only say ``Here, $w(s, p^2)$ are model-independent functions that depend on the collision center-of-mass energy $s$ and the dilepton invariant mass..." \cite{deBlas:2012qp}.  If we \emph{assume} that they are the same\footnote{Manuel Perez-Victoria informed us \cite{Perez-Victoria:2017}  that $w_{u,d}$ in  \cite{deBlas:2012qp} are the same as in \cite{Carena:2004xs}.} as in \cite{Carena:2004xs}, the numerical factor in their equation (3.8) is correct. The authors of \cite{deBlas:2012qp} do not comment on the discrepancy between their equation (3.8) and equation (3.8) of \cite{Carena:2004xs}. 

Similarly, there are expressions in the literature in which the correct differential $Z^\prime$ cross section is given, but not in the exact form of our equation (\ref{sigma6}),  see \cite{Langacker:1984dc, Han:2013mra}. One can, in principle, derive our equation (\ref{sigma6})  from them. In practice, the wrong equation is the one that was used in \cite{Carena:2004xs}, \cite{Accomando:2010fz}, and \cite{Patrignani:2016xqp}.

\subsection{Summary}

Perhaps because of the non-standard definition of the $Z$ boson coupling in (\ref{Hamberg_gz}), the results of  \cite{Hamberg:1990np} seem to have been transcribed incorrectly to the case of a $Z^\prime$ in \cite{Carena:2004xs}. In any case, the quoted result for the cross section is a factor 8 too small. How has that affected $Z^\prime$ searches? 

\section{Effect on experimental searches}\label{experiment}

\subsection{Implementation in Pythia}
Experimental searches for $Z^\prime$ resonances use Pythia, so the implementation of $Z^\prime$ models in Pythia is important. To the best of our knowledge, the first Pythia documentation to discuss $Z^\prime$ is \cite{Sjostrand:1995iq}. It predates \cite{Carena:2004xs} and we assume that it uses the SM $Z$ calculations. See also \cite{Ciobanu:2005pv}  for a detailed discussion of  $Z^\prime$ models implementation in Pythia.

\subsection{$Z^\prime$ constraints in reference \cite{Carena:2004xs}}
Reference \cite{Carena:2004xs} has used preliminary unpublished data from the CDF and D0 experiments to present the first $c_u-c_d$  exclusion plot.  See references 26 and 27 in \cite{Carena:2004xs}. In particular they presented an exclusion plot using the CDF data presented at SUSY 2004. This talk is not available on the conference website \cite{SUSY2004}. The proceedings contribution that is available on the website does not include the data \cite{SUSY2004}. Therefore we cannot check the plot presented in  \cite{Carena:2004xs}. 

\subsection{$Z^\prime$ constraints in reference \cite{Accomando:2010fz}}\label{Accomando}
In 2010 reference \cite{Accomando:2010fz} presented a detailed study of the prospects for setting limits on $Z^\prime$ using early LHC data  \cite{Accomando:2010fz}.  In particular they have used D0 data published in \cite{Abazov:2010ti} to present a $c_u-c_d$  exclusion plot in figure 7 of \cite{Accomando:2010fz}. 

As \cite{Carena:2004xs}, they list the wrong expression for the cross section in equation 3.7 of the journal version of \cite{Accomando:2010fz} (equation II.8 in the arXiv.org version).  On the other hand, they say in section III.B of the journal version of \cite{Accomando:2010fz} (section II.C in the arXiv.org version): ``In the following, we take into account QCD NNLO effects as implemented in the WZPROD program [54-56]...We have adopted this package for simulating the $Z^\prime$ production, and have linked it to an updated set of parton density functions (PDFs)." The references they cite ([54-56] above) include \cite{Hamberg:1990np}. If adapted correctly, they might have the correct expression for the cross section.    

In order to decide which is the case, we created our own $c_u-c_d$  exclusion plot  based on the D0 data  published in \cite{Abazov:2010ti}. We use the ``LO" MSTW2008 PDFs \cite{Martin:2009iq} obtained from \cite{MSTW2008}.  To match  \cite{Accomando:2010fz} as closely as possible, we include  their $K_{\mbox{\scriptsize NNLO}}$ factors in  $\sigma^{\mbox{\scriptsize NNLO}}_{\ell^+\ell^-}\simeq K_{\mbox{\scriptsize NNLO}}^{\mbox{\scriptsize }} \,\sigma^{\mbox{\scriptsize LO}}_{\ell^+\ell^-}\,$ to set the limits. The values of $K_{\mbox{\scriptsize NNLO}}$ are listed in the appendix of \cite{Accomando:2010fz}. We use Table III corresponding to MSTW 2008. 

\begin{figure}
\centering
\includegraphics[height=8cm,keepaspectratio,trim=0cm -3.5cm 0  0cm ]{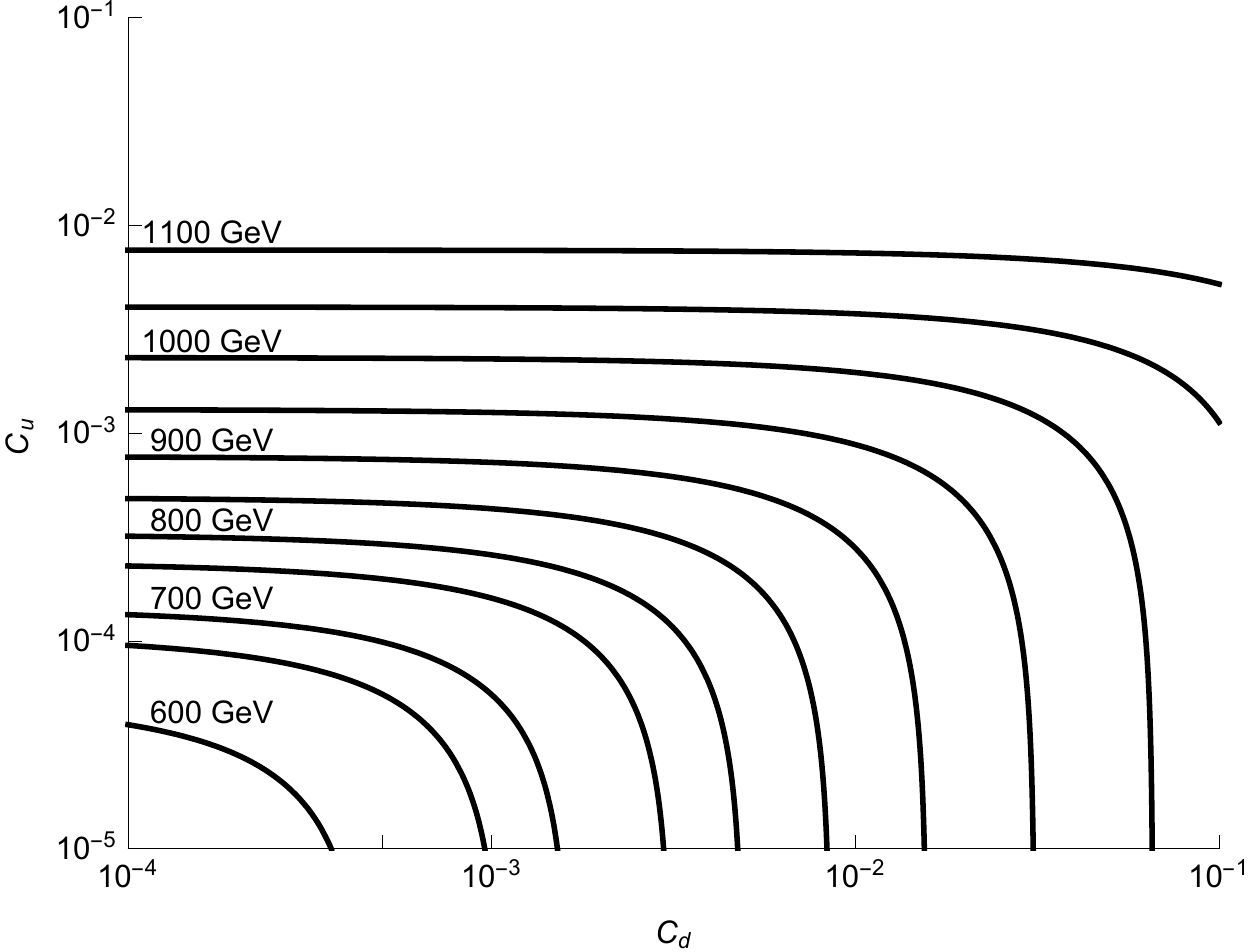}
\hspace{1cm}
\includegraphics[height=9cm,keepaspectratio ]{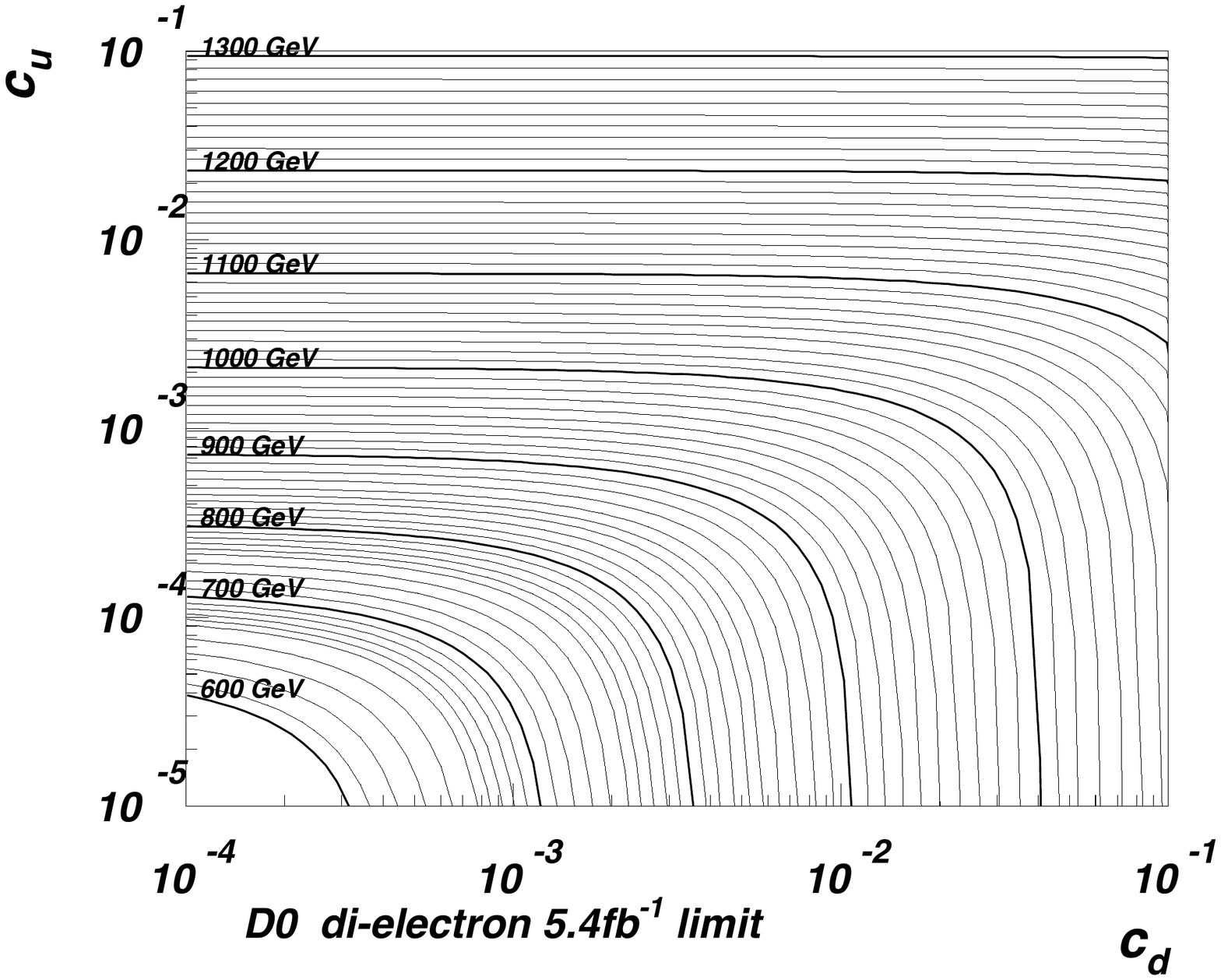}
\vspace{-1.75cm}
\caption{\label{fig1} Left hand side: Our $c_u-c_d$ exclusion plot based on the D0 data \cite{Abazov:2010ti} and equation (\ref{sigma6}). Right hand side: Reference \cite{Accomando:2010fz} $c_u-c_d$ exclusion plot based on the D0 data \cite{Abazov:2010ti} and equation (\ref{sigma48}). }
\end{figure}

Our results are presented in figure \ref{fig1}. On the left hand side of figure \ref{fig1} we show the $c_u-c_d$ contours corresponding to $Z^\prime$ masses of 600 GeV-1100 GeV in 50 GeV increments, as reported by D0. On the right hand side we show the   corresponding plot from reference \cite{Accomando:2010fz}, taken from the right hand side of Figure 7 of  \cite{Accomando:2010fz}. Notice that \cite{Accomando:2010fz} has also \emph{extrapolated} the D0 data to higher masses. The original D0 data only report exclusions up to 1100 GeV.  Although not stated explicitly, \cite{Accomando:2010fz} has presumably interpolated D0 data to generate a much denser plot.  

Despite these, it is easy to see that the two exclusion plots agree in the 600 GeV-1100 GeV range. This implies that despite claiming to have used equation (\ref{sigma48}), \cite{Accomando:2010fz} got the correct exclusion plot, presumably by using \cite{Hamberg:1990np}.  We have checked that using equation (\ref{sigma48}) and/or omitting $K_{\mbox{\scriptsize NNLO}}$  gives a plot that is different from the one presented in \cite{Accomando:2010fz}.

 \subsection{LHC $Z^\prime$ constraints}

To the best of our knowledge, only CMS has produced  $c_u-c_d$ exclusion plots for LHC data. Such plots were given for 7 TeV data in \cite{Chatrchyan:2011wq,Chatrchyan:2012it} and 8 TeV data in \cite{Khachatryan:2014fba} LHC data. Unlike the D0 data \cite{Abazov:2010ti} that included limits on the cross section, CMS has only  produced exclusion plots. As a result we cannot check whether they are using the correct expression for the cross section. Since these CMS analyses \cite{Chatrchyan:2011wq,Chatrchyan:2012it, Khachatryan:2014fba} cite  \cite{Carena:2004xs}  and \cite{Accomando:2010fz} as the theoretical basis, CMS should check the effect of the wrong factor in  (\ref{sigma48}) on their limits.

\section{Conclusions}\label{conclusions}
A heavy $U(1)'$ gauge boson, commonly referred to as a $Z^\prime$, is predicted in many BSM models. If it decays to leptons, such a $Z^\prime$ can appear very clearly in the di-lepton invariant mass spectrum. So far experimental searches have not found such a signal and only limits on the cross section were set. If such experimental data are properly presented it can be very useful in constraining and improving $Z^\prime$ BSM models.

The most useful way to present such data was suggested in  \cite{Carena:2004xs}, where the cross section is ``factorized" into a sum over products of $Z^\prime$ model-dependent $c_i$ and $Z^\prime$ model-independent $w_i$ that depend on the PDFs. Since the latter can be calculated, the cross section can be translated into constraints on the model-dependent $c_i$ without the need to resort to specific models. This method was used to analyze Tevatron data in  \cite{Carena:2004xs} and \cite{Accomando:2010fz} and LHC data in  \cite{Chatrchyan:2011wq,Chatrchyan:2012it, Khachatryan:2014fba}.

We have repeated the derivation of  \cite{Carena:2004xs} in section \ref{theory} and we find that the expression given there is 8 times too small.  Thus instead of equation (\ref{sigma48}) used by  \cite{Carena:2004xs}, one should use equation (\ref{sigma6}) presented here.  We confirmed our calculations by comparing them to the known cases of the photon \cite{Halzen:1984mc} and the SM $Z$ \cite{Hamberg:1990np} Drell-Yan. It should be noted that  \cite{Hamberg:1990np} uses an unusual  normalization for the SM $Z$ coupling, and one should be careful in generalizing  \cite{Hamberg:1990np} to the case of a $Z^\prime$. 

Taken at face value, our result  might imply that the experimental constraints that use the wrong formula are off by a factor of 8. To test that, we have tried to check the existing exclusion plots in section \ref{experiment}. The original paper \cite{Carena:2004xs} has used CDF data  that was only shown at a conference \cite{SUSY2004} and is not available online or in the proceedings. As a result,  we cannot check the exclusion plot in \cite{Carena:2004xs}.  A later paper,  \cite{Accomando:2010fz},  has claimed to use the (\ref{sigma48}) and D0 data \cite{Abazov:2010ti} to produce an exclusion plot. In this case we can redo their analysis. As figure \ref{fig1} shows, the exclusion plot of  \cite{Accomando:2010fz} agrees with our result. This implies that they have not actually used  (\ref{sigma48}) in their analysis.  Finally, for the case of LHC data, exclusion plots were produced by CMS in \cite{Chatrchyan:2011wq,Chatrchyan:2012it, Khachatryan:2014fba}. These do not contain enough information that will allow us to reproduce them. Therefore we cannot determine whether they are correct.

In summary, considering the important implications, we encourage the high energy physics community to re-examine past analyses that have used the wrong expression for the cross section.

\emph {Note added.} After this paper was completed, Alexander Belyaev  informed us \cite{Belyaev:2017} that both  \cite{Accomando:2010fz} and the CMS analyses \cite{Chatrchyan:2011wq,Chatrchyan:2012it, Khachatryan:2014fba}  may list the wrong expression (\ref{sigma48}), but not actually use it.  The coefficients in front of $c_u$ and $c_d$ in the cross section were found numerically for all of these papers. This confirms our observation in section \ref{Accomando} and implies that the exclusion plots in  \cite{Chatrchyan:2011wq,Chatrchyan:2012it, Khachatryan:2014fba}  are valid.

\vskip 0.2in
\noindent
{\bf Acknowledgements}
\vskip 0.1in
\noindent
 This work was supported by Department of Energy Grant No. DE-SC0007983.

\end{document}